\documentclass[aps,preprint,amssymb,12pt,floatfix]{revtex4}
\usepackage{subfigure}
\usepackage[pdftex]{graphicx}
\usepackage{lscape,graphicx}
\usepackage{rotating}
\usepackage{epstopdf}
\usepackage{color}
\usepackage{amsmath,amsthm}

\begin{document}
\title{Ripping RNA by Force using Gaussian Network Models}
\author{Changbong Hyeon}
\affiliation{School of Computational Sciences, Korea Institute for Advanced Study, Seoul 130-722, Republic of Korea}
\author{D. Thirumalai}
\affiliation{Department of Chemistry, University of Texas at Austin, TX 78712, USA}
\date{\today}
\begin{abstract}
Using force as a probe to map the folding landscapes of RNA molecules has become a reality thanks to major advances in single molecule pulling experiments. Although the unfolding pathways under tension are complicated to predict studies in the context of proteins have shown that topology plays is the major determinant of the unfolding landscapes. By building on this finding we study the responses of RNA molecules to force by adapting Gaussian network model (GNM) that represents RNAs using a bead-spring network with isotropic interactions. 
Cross-correlation matrices of residue fluctuations, which are analytically calculated using GNM even upon application of mechanical force, show distinct allosteric communication as RNAs rupture. 
The model is used to calculate the  force-extension curves at full thermodynamic equilibrium, and the corresponding 
 unfolding pathways of four RNA molecules subject to a quasi-statically increased force. 
Our study finds that the analysis using GNM captures qualitatively the unfolding pathway of \emph{T}. ribozyme elucidated by the optical tweezers measurement. However, the simple model is not sufficient to capture subtle features, such as bifurcation in the unfolding pathways or the ion effects, in the forced-unfolding of RNAs.
\end{abstract}
\maketitle

\section{Introduction}
While rigid body dynamics under mechanical stress is  well studied in classical mechanics and engineering \cite{landau1986theory}, 
similar considerations in the context of biopolymers  began in earnest only recently with the advent of single molecule (SM) techniques. These methods are now routinely used   to manipulate the biopolymers \cite{Evans97BJ,Strick1996Science,Bustamante01Sci,Onoa2003Science,Block06Science,Greenleaf08Science,Dittmore2011PRL}. 
The renewed interest in the semiflexible chain \cite{FreyPRL96,HaJCP95,HaJCP97} is partly due to the stretching experiments on dsDNA and multidomain proteins whose force-extension curves are well characterized by the worm-like chain model \cite{Bustamante94SCI}. 
SM manipulation techniques using the atomic force microscopy (AFM) and laser optical tweezers (LOT) have yielded great insights into the elasticity of nucleic acids as well as quantitive glimpses into the folding landscapes of proteins and RNA and their complexes.  
In particular, the use of mechanical force provides a viable means to measure and decipher the underlying characteristics of the free energy landscape, such as the position of transition state, ruggedness, and the height of free energy barrier \cite{Bustamante01Sci,TinocoARBBS04,Hyeon03PNAS,Block06Science,Hyeon07JP,Hyeon08PNAS,hinczewski2013PNAS}, and the extent of dynamic/static disorder in the pulled molecules \cite{Hyeon14PRL,Hinczewski2016PNAS}. 
Quantitative understandings into structure and energetics of DNA \cite{PrentissPNAS03,Block06Science}, RNA \cite{Bustamante01Sci,Onoa2003Science,Greenleaf08Science}, and proteins \cite{GaubSCI97,Mickler07PNAS,FernandezNature99,Fernandez04Science,stigler2011complex,zhuravlev2016PNAS,Klimov99PNAS}, which are difficult to decipher using conventional ensemble experiments, have become available by decoding the origin of molecular responses under tension. 

Thanks to major advances in SM force measurement, developments in 
theories, computations, and experiments have cross-fertilized to produce comprehensive understanding of biomolecules on nanometer scales.  
In particular, steered molecular dynamics simulations pioneered by Schulten and coworkers made a major contribution to the field at the early stage of the development \cite{SchultenBJ98,Schulten}. These studies, which have become the standard tool for experimentalists, have been used to visualize the sequence of unfolding events that might occur during the unraveling of proteins under tension at atomic resolution. 
However, the time scale gap between the computational approach using atomically detailed model and experiment has not been resolved  despite progress in computer technologies.   In a practically limited computation time, large forces and loading rates (or pulling speeds) are typically used to accelerate the unbinding or unfolding processes. It is likely that the use of artificially large forces inevitably alters the unfolding pathways from those elucidated from force experiments \cite{Hyeon06Structure}. 
The currently accessible time scale accessible in atomistic detailed simulations is $\lesssim \mathcal{O}(1)$ ms \cite{Shaw10Science}. This restrains the minimum loading rate that can be reached by computer simulation to at best $10^4$ pN/s if one were to study the forced unfolding of a simple RNA hairpin, which reversibly unfolds and folds at $f=10-20$ pN at $r_f\approx 1$ pN/s \cite{Bustamante01Sci}. 
This is four orders of magnitude faster than the experimental loading rate $\sim 1$ pN/s used in LOT. 
The force ramped at rates faster than the relaxation process of the molecules confines their conformations to a narrow phase space. Therefore, it is necessary to explore alternate computational approaches to probe forced-unfolding. 

A decade ago, in an effort to study forced-unfolding dynamics of RNA and proteins at time scales compatible to those in force measurements, 
we proposed a new class of minimal coarse-grained models, called the self-organized (SOP) polymer model \cite{Hyeon06Structure,HyeonBJ07,zhmurov2011Structure}. 
The premise underlying the creation of the SOP model is  that unfolding pathways of biomolecule is mainly determined by the overall network of inter-residue interactions in the native state, which are disrupted stochastically under a given perturbation, and the chain connectivity along the polymer backbone. 
The SOP model allowed us to simulate small-to-medium sized biomolecules at loading rates comparable to the real experiments \cite{HyeonBJ07,Hyeon06Structure}. In the process, we demonstrated that the unfolding pathways of biomolecule are explicitly dependent on the pulling speed \cite{Hyeon06Structure}. 
Furthermore, we revealed that the unfolding pathways of GFP bifurcate into two alternative pathways (opening of GFP barrel from either N- or C- terminus), which was difficult to capture by analyzing force-extension curves from AFM pulling experiments alone \cite{Mickler07PNAS}. These predictions were not only unanticipated in experiments but are difficult to predict using standard molecular dynamics simulations.


The Gaussian Network Model (GNM) \cite{BaharPRL97} (or elastic network model (ENM) \cite{TirionPRL96,BaharBJ01}), used to describe conformational fluctuations and intra-molecular allosteric communications of biomolecules in the native states \cite{BaharMSB06,Zheng05Structure,Zheng06PNAS,Bahar05COSB,BrooksPNAS00}, provides an alternative way to predict the force-induced rupture of RNA and proteins. 
Conformational fluctuations of proteins in their native state is faithfully described by low-frequency normal modes \cite{Zheng06PNAS}, which can be captured by using the topological information in the native contact maps. 
The low frequency modes remain insensitive to the details of interaction, such as strength and nature of interactions (vdW, H-bond, electrostatic). 
This observation justifies the use of coarse-grained representations of molecular systems for simulating at large length and time scales.  
Indeed, GNM is excellent in reproducing the relative amplitude of residue fluctuation, which is often compared with Debye-Waller factor.

Here, adapting the study of Srivastava and Graneck, who considered thermal and force-induced unfolding of proteins by extending the formalism of GNM \cite{Srivastava13PRL}, 
we explore force-induced rupture of RNA molecules. 
In the framework of GNM, the dynamic cross-correlation between residue fluctuations at finite temperature is related to the inverse Kirchhoff matrix, containing information on network connectivity. The response of the system under force is related to squared inverse Kirchhoff matrix.  
Mechanical force can, in principle, be applied to any site in the structure, revealing the allosteric communication spread throughout the structure. 
In addition, the force-induced unfolding experiment is mimicked by applying forces on the two ends (or two distinct sites). 
Increasing forces induces a large fluctuation in the non-covalently bonded residue pairs ($|i-j|\ge 2$) and disrupts them, which can be implemented by modifying the connectivity matrix at each value of the applied force. 
We show that GNM using topological information alone can satisfactorily unveil the forced-unfolding pathway of RNA molecule at equilibrium. \\

\section{Theory}
{\bf Gaussian network model under force:} One of the aims of using GNM \cite{BaharPRL97} is to elucidate the natural conformational dynamics arising solely from the native state contact map. 
The reasonable agreement between theoretically computed mean square fluctuation (MSF) and the crystallographic B-factor suggests that residue-specific interactions are of secondary importance in characterizing the dynamics of a biomolecule in the native state \cite{BaharPRL97}. 
This is further substantiated by a number of studies that show fluctuations and correlated motions of the native conformation 
are  satisfactorily described solely using the information on native contacts. 

In GNM, vectors linking residue pairs in contact (${\bf R}_{ij}\equiv {\bf R}_i-{\bf R}_j$)
relative to those in the native state ${\bf R}^o_{ij}\equiv {\bf R}^o_i-{\bf R}^o_j$ are harmonically restrained to have the following form, 
\begin{align}
\mathcal{H}_0=\frac{\gamma}{2}\sum_{(i,j)}({\bf R}_{ij}-{\bf R}_{ij}^o)^2=\frac{\gamma}{2}\sum_{(i,j)}(\delta{\bf R}_i-\delta{\bf R}_j)^2.
\end{align}
where $(i,j)$ denotes the list of residue pairs in contact satisfying the geometrical condition $|{\bf R}_{ij}^o|<R_c$ in the native state. 
The native state molecular coordinates are obtained from the structures in the Protein Data Bank (PDB). Let 
$\delta{\bf R}_i\equiv {\bf R}_i-{\bf R}_i^o$ be the deviation of the $i$-th bead from its coordinate in the native state.    
This Hamiltonian given above can be written as $\mathcal{H}_0=\frac{\gamma}{2}\delta{\bf R}^T\cdot{\bf \Gamma}\cdot\delta {\bf R}$ where 
$\delta {\bf R}=(\delta {\bf R}_1,\delta{\bf R}_2,\ldots,\delta{\bf R}_N)^T$. 
Although one could consider an improvement of the model, GNM \cite{BaharPRL97} merely uses a uniform bond strength $\gamma$ regardless of the type of the bond.  This is sufficient to illustrate the essential ideas.
The Laplacian (Kirchhoff) matrix ${\bf \Gamma}$ where $\Gamma_{ij}$ results from the second derivative of harmonic potential with respect to $\delta {\bf R}_i$ and $\delta {\bf R}_j$. 
Note that the residue motion is ``isotropic" in GNM, which  reduces  
$3N-6$ degrees of freedom to $N-1$. 
The cross-correlation between the $i$-th and $j$-th residues is obtained from the partition function $Z_N=\int \mathcal{D}[\delta{\bf R}]e^{-\beta \mathcal{H}_0}=(2\pi k_BT/\gamma)^{3(N-1)/2}$ as, 
\begin{align}
\langle\delta{\bf R}_i\cdot\delta{\bf R}_j\rangle = -\frac{2k_BT}{\gamma}\frac{\partial\log Z_N}{\partial \Gamma_{ij}}=\frac{3k_BT}{\gamma}\left({\bf \Gamma}^{-1}\right)_{ij}.  
\end{align} 
The inverse Kirchhoff matrix can be computed by decomposing it into eigenmodes using  ${\bf \Gamma}^{-1} = {\bf U}{\bf \Lambda}^{-1}{\bf U}^{-1} = {\bf U}{\bf \Lambda}^{-1}{\bf U}^{T}=\sum_{k=2}^N\lambda_k^{-1}{\bf u}_k{\bf u}_k^T$ with $\lambda_k$ and ${\bf u}_k$ being the k-th eigenvalue and eigenvector, respectively. 
Note that the lowest mode ($k=1$), corresponding to the overall translation, is removed from the sum. 
In the absence of force ($f=0$), the self-correlation (or mean square fluctuation, MSF) $\langle\delta {\bf R}^2_i\rangle$ is compared with the crystallographic (Debye-Waller) temperature factor, $B_i = \frac{8\pi^2}{3}\langle(\delta {\bf R}_i)^2\rangle$, to determine the value of the parameter $\gamma$.  The cut-off distance $R_c$ is also determined by appealing to experimental B-factors.
An optimum choice of cut-off distance $R_c$, which minimizes the difference between the temperature factor and self-correlation is expected. 
For RNA, when the center of mass in each nucleotide is set to be the position of each nucleotide, $R_c \sim (10-20)$ \AA\ reproduces  reasonable results for B-factor with 
$\gamma\sim 10$ pN$\cdot$nm/\AA$^2$. 
For all the RNAs tested here, we chose $R_c = 15$ \AA. 
This choice of $R_c$ also provides better sensitivity to the system in the  bond rupture process. 
When $R_c = 15$ \AA\ is used, we find that the difference between $B_i$ and $8\pi^2/3\times\langle(\delta {\bf R}_i)^2\rangle$ for P4P6 domain (PDB code: 1gid) is minimized with $\gamma\approx  7.4$ pN$\cdot$nm/\AA$^2$. 
\\

{\bf Allosteric communication, inter-residue cross-correlation under tension.} 
When the $k$-th residue of the molecule is perturbed by mechanical force. 
the Hamiltonian becomes 
\begin{align}
\mathcal{H} = \frac{\gamma}{2}\delta {\bf R}^T \cdot {\bf \Gamma}\cdot \delta {\bf R} - {\bf f}_k^T\cdot \delta {\bf R}.
\end{align} 
${\bf f}_k$ is the general force vector acting on the $k$-th residue; thus ${\bf f}_k = f{\bf e_k}$. 
The cross-correlation between the $i$-th and $j$-th residues under mechanical perturbation at $k$-th residue is obtained from the partition function,
\begin{align}
Z_N({\bf f}_k)=\int \mathcal{D}[\delta{\bf R}]e^{-\beta \mathcal{H}}=\left(\frac{2\pi k_BT}{\gamma}\right)^{3(N-1)/2}\left[\det{\bf \Gamma}\right]^{-3/2}\exp\left(\frac{{\bf f}^T_k\cdot{\bf \Gamma}^{-1}\cdot {\bf f}_k}{2\gamma k_BT}\right)
\end{align}
as follows: 
\begin{align}
\langle\delta{\bf R}_i\cdot\delta{\bf R}_j\rangle &=-\frac{2k_BT}{\gamma}\frac{\partial\log Z_N({\bf f}_k)}{\partial \Gamma_{ij}}\nonumber\\
&={\bf T}_{ij}+{\bf F}_{ij}^k\nonumber\\
&=\frac{3k_BT}{\gamma}\left({\bf \Gamma}^{-1}\right)_{ij}
+\frac{f^2}{\gamma^2}\left({\bf \Gamma}^{-1}\right)_{ki}\left({\bf \Gamma}^{-1}\right)_{jk}
\label{eqn:crosscorr}
\end{align}
Because of the linear coupling between force and displacement in Eq. (3) the fluctuation-fluctuation cross-correlation ($\langle\delta{\bf R}_i\cdot\delta{\bf R}_j\rangle$) is fully decomposed into temperature (${\bf T}_{ij}$) and force (${\bf F}^k_{ij}$) contributions.  
The $f$-dependent term in the Eq.\ref{eqn:crosscorr} results from the following matrix calculus: $\partial ({\bf \Gamma}{\bf \Gamma}^{-1})/\partial \Gamma_{ij}= 0$ 
and $\partial {\bf \Gamma}/\partial \Gamma_{ij}={\bf e}_i{\bf e}_j^T$,
thus, $\partial {\bf \Gamma}^{-1}/\partial \Gamma_{ij} = -{\bf \Gamma}^{-1}{\bf e}_i{\bf e}_j^T{\bf \Gamma}^{-1}$. 
It is noteworthy that the thermal fluctuations (${\bf T}_{ij}$) are decoupled from the effect of the applied perturbation. 
The allosteric cross-correlation between the two distinct parts of the structure due to mechanical perturbation on $k$-th residue is gleaned by inspecting ${\bf F}^k_{ij}$. 
 \\

{\bf Force-extension curves:} 
When tension is applied to the two ends of a molecule at $i=1$ and $N$, the energy Hamiltonian is written as 
\begin{align}
\mathcal{H} = \frac{\gamma}{2}\delta {\bf R}^T \cdot {\bf \Gamma}_f\cdot \delta {\bf R} - ({\bf f}_N^T -{\bf f}_1^T)\cdot \delta {\bf R}.
\end{align} 
and the corresponding partition function and dynamical cross-correlation are: 
\begin{align}
Z_N=\int \mathcal{D}[\delta{\bf R}]e^{-\beta \mathcal{H}}=\left(\frac{2\pi k_BT}{\gamma}\right)^{3(N-1)/2}\left[\det{\bf \Gamma}\right]^{-3/2}\exp\left(\frac{({\bf f}^T_N-{\bf f}^T_1)\cdot{\bf \Gamma}_f^{-1}\cdot ({\bf f}_N-{\bf f}_1)}{2\gamma k_BT}\right)
\end{align}
and 
\begin{align}
\langle\delta{\bf R}_i\cdot\delta{\bf R}_j\rangle &=-\frac{2k_BT}{\gamma}\frac{\partial\log Z_N}{\partial (\Gamma_{f})_{ij}}\nonumber\\
&={\bf T}_{ij}+{\bf F}_{ij}^{N,1}\nonumber\\
&=\frac{3k_BT}{\gamma}\left({\bf \Gamma}_f^{-1}\right)_{ij}
+\frac{f^2}{\gamma^2}[\left({\bf \Gamma}_f^{-1}\right)_{Ni}-\left({\bf \Gamma}_f^{-1}\right)_{1i}][\left({\bf \Gamma}_f^{-1}\right)_{jN}-\left({\bf \Gamma}_f^{-1}\right)_{j1}]. 
\label{eqn:crosscorr_f}
\end{align}
Here, we make the $f$-dependence of Kirchhoff matrix explicit by putting a subscript, which becomes clear in the algorithm used to calculate force-extension curves (see below). 

The equilibrium force-extension curve is computed using Eq.\ref{eqn:crosscorr_f}. 
For the residue pairs making bond, say a pair between $i$-th and $j$-th bead, 
mean square fluctuation $\langle (\delta{\bf R}_{ij})^2\rangle\equiv\langle (\delta {\bf R}_i-\delta {\bf R}_j)^2\rangle=\langle(\delta {\bf R}_i)^2\rangle+\langle(\delta {\bf R}_j)^2\rangle-2\langle(\delta {\bf R}_i\cdot\delta {\bf R}_j)\rangle$. 
To be specific, 
\begin{align}
\langle (\delta{\bf R}_{ij})^2\rangle=\langle (\delta{\bf R}_{ij})^2\rangle_T+\langle (\delta{\bf R}_{ij})^2\rangle_f
\end{align}
where 
\begin{align}
\langle (\delta{\bf R}_{ij})^2\rangle_T=\frac{3k_BT}{\gamma}[({\bf \Gamma}_f^{-1})_{ii}+({\bf \Gamma}_f^{-1})_{jj}-2({\bf \Gamma}_f^{-1})_{ij}]\nonumber
\end{align}
\begin{align}
\langle (\delta{\bf R}_{ij})^2\rangle_f
&=\frac{f^2}{\gamma^2}[\left({\bf \Gamma}_f^{-1}\right)_{Ni}-\left({\bf \Gamma}_f^{-1}\right)_{1i}]^2\nonumber\\
&+\frac{f^2}{\gamma^2}[\left({\bf \Gamma}_f^{-1}\right)_{Nj}-\left({\bf \Gamma}_f^{-1}\right)_{1j}]^2\nonumber\\
&-2\frac{f^2}{\gamma^2}[\left({\bf \Gamma}_f^{-1}\right)_{Ni}-\left({\bf \Gamma}_f^{-1}\right)_{1i}][\left({\bf \Gamma}_f^{-1}\right)_{jN}-\left({\bf \Gamma}_f^{-1}\right)_{j1}]\nonumber
\end{align}
For a given $f$, we assume that a bound residue pair between $i$ and $j$ is disrupted when 
\begin{align}\phi\equiv\frac{\langle (\delta{\bf R}_{ij})^2\rangle_f}{\langle (\delta{\bf R}_{ij})^2\rangle_T}>1.
\end{align} 
Breakage of the non-covalent bond between $i$ and $j$ pair with $({\bf \Gamma}_f)_{ij}=-1$ and $|i-j|\geq 2$ is implemented in the computer algorithm by modifying the elements of the Kirchhoff matrix as 
\begin{align}
&({\bf\Gamma}_f)_{ij}=-1\longrightarrow 0\nonumber\\
&({\bf \Gamma}_f)_{ji}=-1\longrightarrow 0\nonumber\\
&({\bf \Gamma}_f)_{ii}\longrightarrow ({\bf \Gamma}_f)_{ii}-1\nonumber\\ 
&({\bf \Gamma}_f)_{jj}\longrightarrow ({\bf \Gamma}_f)_{jj}-1. 
\end{align}
The original Kirchhoff matrix is modified at each value of an increasing $f$ whenever the condition $\phi>1$ is identified for any pair of residue until all the non-covalent residue pairs are ripped. 
The extension is computed as the size of free ends using the Kirchhoff matrix, ${\bf \Gamma}_f$. 
Thus, reconstruction of Kirchhoff matrix with increasing force produces force-extension curve. 
We selected $\phi =0.002$, so that the transition force for P4P6 domain is at $f_c\sim$ (10 -- 15) pN.
\\

\section{Results}
We applied the GNM formalism under mechanical perturbation to four RNA molecules whose 3D structures are available: 
(i) 55-nt domain IIa of the Hepatitis C Viral (HCV) genome (PDB code: 1p5m) \cite{PuglisiNSB03}; 
(ii) prohead RNA (PDB code: 1foq) \cite{RossmanNature00};
(iii) P4P6 domain of \emph{T}.ribozyme (PDB code: 1gid) \cite{cate1996Science}; 
(iv) \emph{T}. ribozyme (the atomic coordinates of a theoretically modeled \emph{T}. thermophila ribozyme, TtLSU.pdb, was obtained from the Group I and II sections in the website http://www-ibmc.u-strasbg.fr/upr9002/westhof) \cite{WesthofCB96}. 
We first discuss the effect of mechanical perturbation on the dynamical cross-correlation  using \emph{T}. ribozyme, followed by the calculation of force-extension curves for all four RNAs. 
\\

\begin{figure}
	\centering
	\includegraphics[width=1\linewidth]{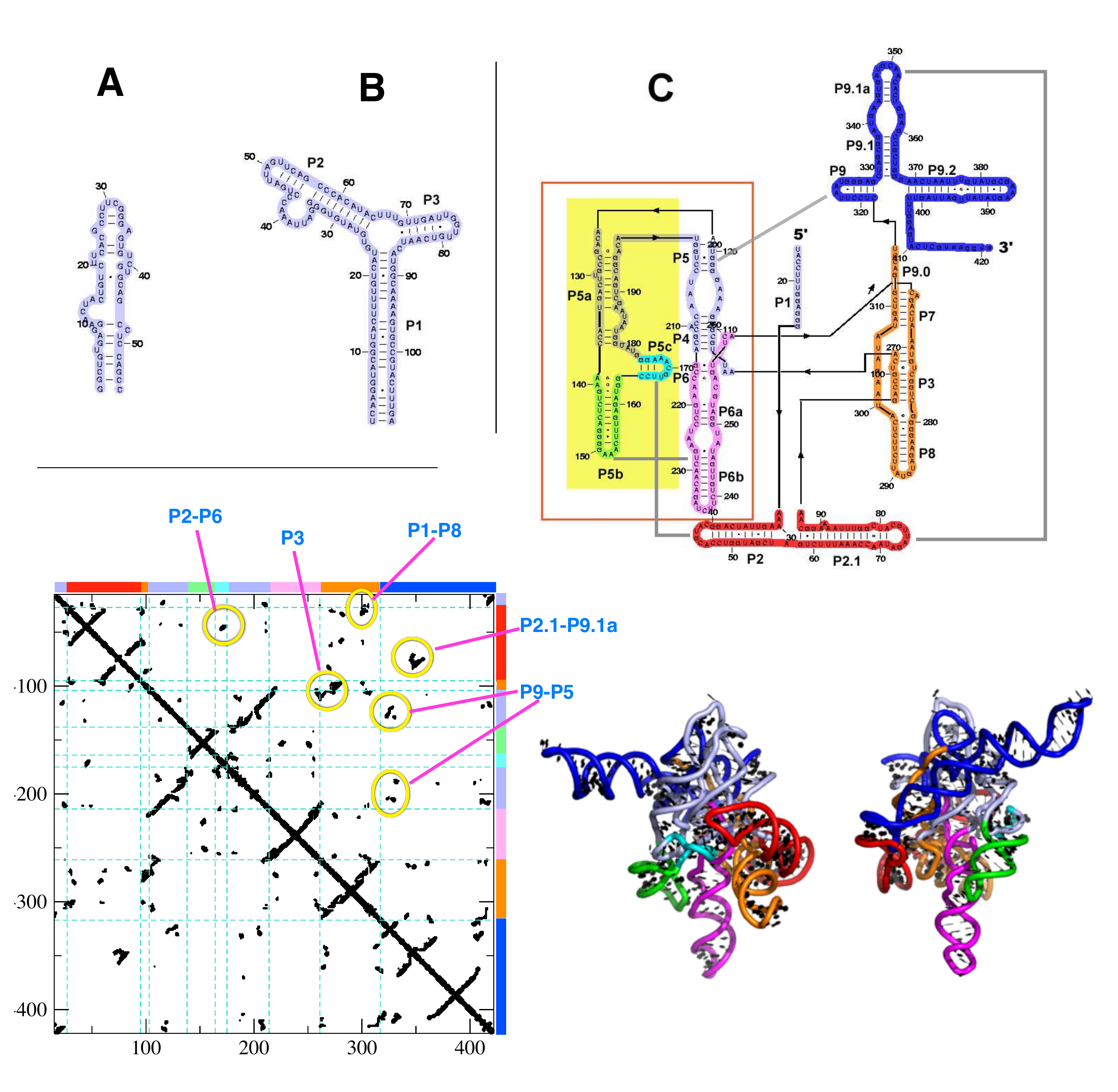}
	\caption{Secondary structures of RNA. 
	(A) IIa domain of HCV IRES RNA (PDB code: 1p5m). 
	(B) prohead RNA (PDB code : 1foq). 
	(C) \emph{Tetrahymena} ribozyme. The region enclosed by a box corresponds to P4P6 domain (PDB code: 1gid). 
	The contact map of \emph{T}. ribozyme and 3D structures are shown at the bottom. 
	Some of the key tertiary contacts are highlighted with yellow circles. 
	}
	\label{fig:structures}
\end{figure}

{\bf Allosteric communication of \emph{T}. ribozyme in the presence of mechanical force:} 
Using the Kirchhoff matrix ${\bf \Gamma}$ of \emph{T}. ribozyme we calculated the temperature-induced dynamical cross-correlation with MSF, $\langle(\delta {\bf R})^2\rangle$ (Fig.\ref{fig:Tij}) and 
the force-induced cross-correlation matrix ${\bf F}^k$ for $k=1,\ldots,N$ without the prefactor $f^2/\gamma^2$ (Fig.\ref{fig:perturbation} and Supporting Movie). 
Some of the results which exhibit dramatic effect of tension on cross-correlation are demonstrated along with the map of secondary structure of \emph{T}. ribozyme (Fig.\ref{fig:perturbation}), and full results of perturbations are provided in the Supporting Movie. 
The extent of response to the perturbation is larger for those residues with large MSF that are not constrained by the rest of the structure. 
A perturbation on P9.2, which is clearly isolated from other parts (see contact map and 3D structure in Fig.\ref{fig:structures}), induces large cross-correlations over the entire structure. 
Strong correlation is found between two sites when they are in direct contact in the structure. 
For example, the perturbation on $k=325$, a part of P9 helix, induces enhanced fluctuation on P5 helix which is in direct contact with P9 via tertiary interactions. 
There is a  strong cross-correlation between P5b and P6b when $k=151$ is perturbed, between P2 and P5c for $k=45$, between P2.1 and P9.1a for $k=74$, between the residues consisting of the internal multiloop formed among P9, P9.1, P9.2 for $k=336$. 
While one may argue that these are trivial outcomes of the contact information shown in Fig.\ref{fig:structures}C, quantitative prediction of the force-induced, site-dependent cross-correlations demonstrated in Fig.\ref{fig:perturbation} is not possible without calculating ${\bf F}_{ij}^k$.  
Furthermore, the \emph{secondary} cross-correlation between P9.2 and P6b, P9.1a and P7 induced by perturbing P2.1 ($k=74$) is also difficult to predict \emph{a priori}. 

In summary, the entire network responds to mechanical stress at a particular site. 
Although force exerted to the network is linear, the response of the network (\emph{T}. ribozyme) is not trivial, but depends on the network structure. 
The presence of secondary cross-correlation is viewed as a realization of allosteric communication over the entire molecular architecture.  
\\

{\bf Equilibrium force-extension curve:} 
We computed the equilibrium force-extension curves of four RNA molecules. 
All the force-extension curves exhibit a single sharp transition around 10--15 pN, whose value can be adjusted by changing the value of $\phi$. 
Unlike the FECs observed in LOT measurements or simulations which use a finite pulling speed, the details of domain-by-domain unraveling are not clear enough to extract from FEC. 
This is because the FEC from GNM is by nature an outcome at full thermodynamic equilibrium, which corresponds to the situation when the molecule is pulled reversibly using a slow pulling speed \cite{Bustamante02Science}. 
Even if there is a possibility  of unfolding bifurcation, as in our previous Brownian dynamics simulation study of the forced-unfolding of \emph{T}. ribozyme using the SOP model \cite{Hyeon06Structure} it cannot be extracted from a purely elastic network model. If  time is long enough to allow the \emph{T}. ribozyme to fully sample the distinct configurations along the bifurcating pathways, the molecular extension is effectively a value averaged over such stochastic variations of individual trajectories. 

While FEC from GNM in itself is of little use to decode the order of unfolding events, the number of native contacts in each residue, which can be read out from the diagonal elements of the Kirchhoff matrix ($Q_i=({\bf \Gamma}_f)_{ii}$) at each $f$, provides us with information on how each residue loses its contact shared with others as $f$ increases. 
\\

\begin{figure}
	\centering
	\includegraphics[width=1\linewidth]{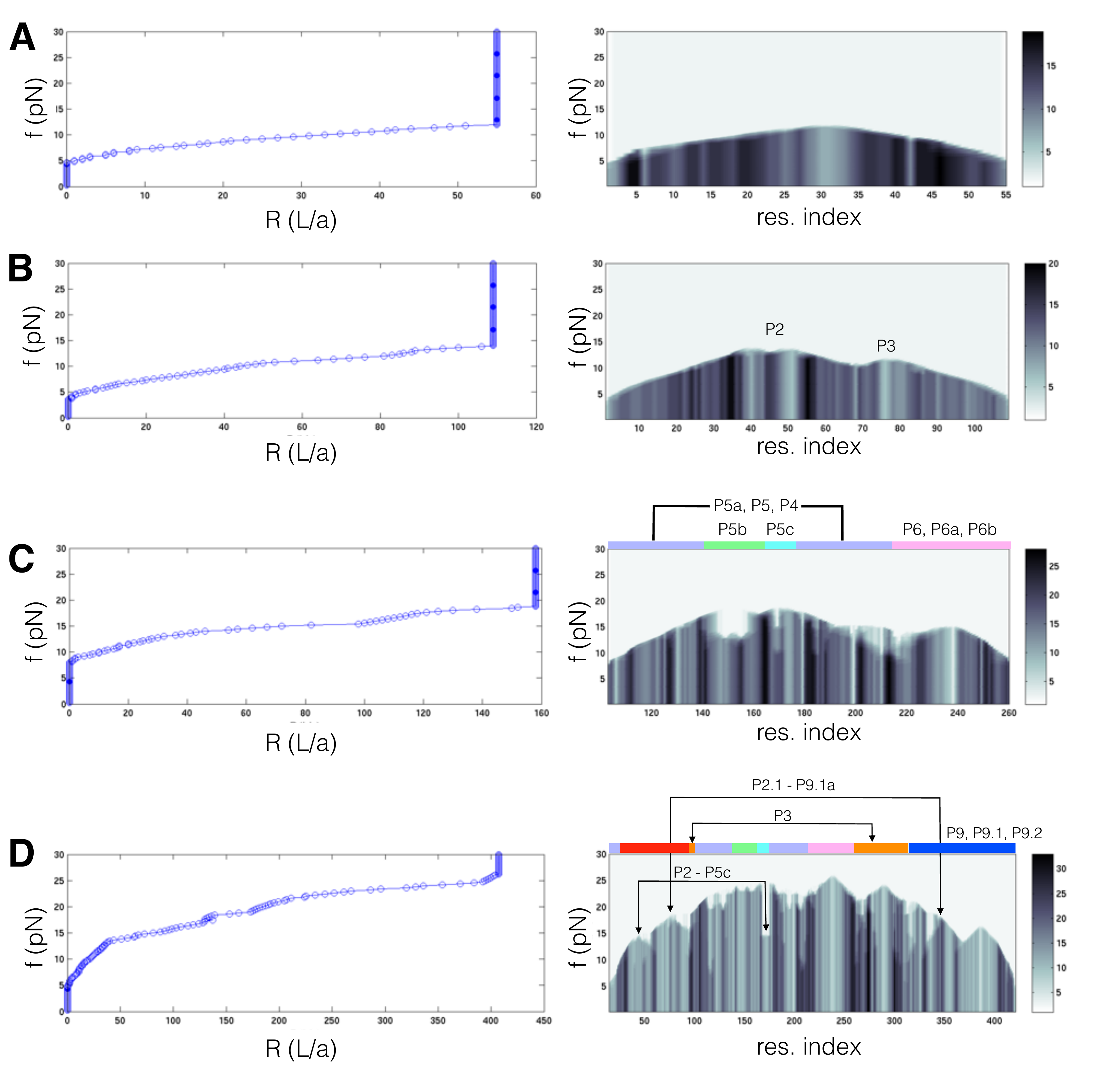}
	\caption{Force-extension curves and number of native contacts ($Q_i=({\bf \Gamma}_f)_{ii}$, diagonal elements of Kirchhoff matrix) with increasing force calculated for 
	(A) IIa domain of HCV IRES RNA (PDB code: 1p5m),  
	(B) prohead RNA (PDB code : 1foq), 
	(C) P4P6 domain, and 
	(D) \emph{Tetrahymena} ribozyme. 
	}
	\label{fig:FEC}
\end{figure}

{\it IIa domain of HCV IRES RNA:} 
Both force-extension curve and history of contact disruption (Fig.\ref{fig:FEC}A) indicate that this hairpin with several bulges unfolds in an all-or-none fashion.  
However, our previous study using Brownian dynamics simulation of SOP model on this RNA 
has revealed two intermediate states between native and fully unfolded states \cite{HyeonBJ07}. 
The absence of signature of intermediate states  is due to the inability of using finite loading rate in the GNM in contrast to the BD simulation used to probe forced-unfolding of IIa domain \cite{HyeonBJ07}. 
 \\
 
{\it Prohead RNA (Two way junction):} 
This two way junctions unfolds via [P1]$\rightarrow$[P3]$\rightarrow$[P2] (Fig.\ref{fig:FEC}B), which is in full accord with the our previous BD simulation result using SOP model at finite loading rate \cite{HyeonBJ07}. 
\\

{\it P4P6 domain:} 
[P6,P6a,P6b] unfolds after the tertiary contacts between P5b and P6b is disrupted, which is followed by the unfolding of P5 helix, and P5abc domain unfolds at the final stage. 
This agrees well with LOT experiment \cite{Onoa2003Science}. 
\\

{\it T. ribozyme:} 
Like other RNA, the unfolding is initiated from the two ends of the ribozyme which composing the peripheral domains [P1], [P2,P2.1], and [P9,P9.1,P9.2], and bond disruptions proceed into the core regions including [P4,P5,P6] and [P3,P7,P8]. 
The evolution profile of bond rupture also captures the early events of tertiary contact disruptions, such as the breakages of P2-P5c and P2.1-P9.1a kissing loop pairs and P3 helix before the disruptions of core helices occur. 
Of note, in our previous work \cite{Hyeon06Structure} on the mechanical unfolding simulations of T. ribozyme using SOP model, 
we found two alternative pathways: 
(1) 
[N]$\rightarrow$[P9.2]$\rightarrow$[P9.1, P9, P9.1a]$\rightarrow$[P2]$\rightarrow$[P2.1]$\rightarrow$[P3, P7, P8]$\rightarrow$[P6]$\rightarrow$[P4, P5]$\rightarrow$[P5a, P5b, P5c];  
(2) [N]$\rightarrow$[P2]$\rightarrow$[P2.1]$\rightarrow$[P9.2]$\rightarrow$[P9, P9.1, P9.1a]$\rightarrow$[P3, P7, P8]$\rightarrow$[P6]$\rightarrow$[P4, P5]$\rightarrow$[P5a, P5b, P5c].   
The first pathway is compatible with the experimentally inferred pathway is [N]$\rightarrow$[P9.2]$\rightarrow$[P9.1]$\rightarrow$[P9, P9.1a]$\rightarrow$[P2, P2.1]$\rightarrow$[P3, P7, P8]$\rightarrow$[P6, P4]$\rightarrow$[P5]$\rightarrow$[P5a, P5b, P5c].
Given the simplicity of the method, the general agreement of the unfolding pathway predicted by GNM with that from the LOT experiment \cite{Onoa2003Science} is remarkable. 
\\

\section{Concluding Remarks}
\begin{figure}
	\centering
	\includegraphics[width=0.5\linewidth]{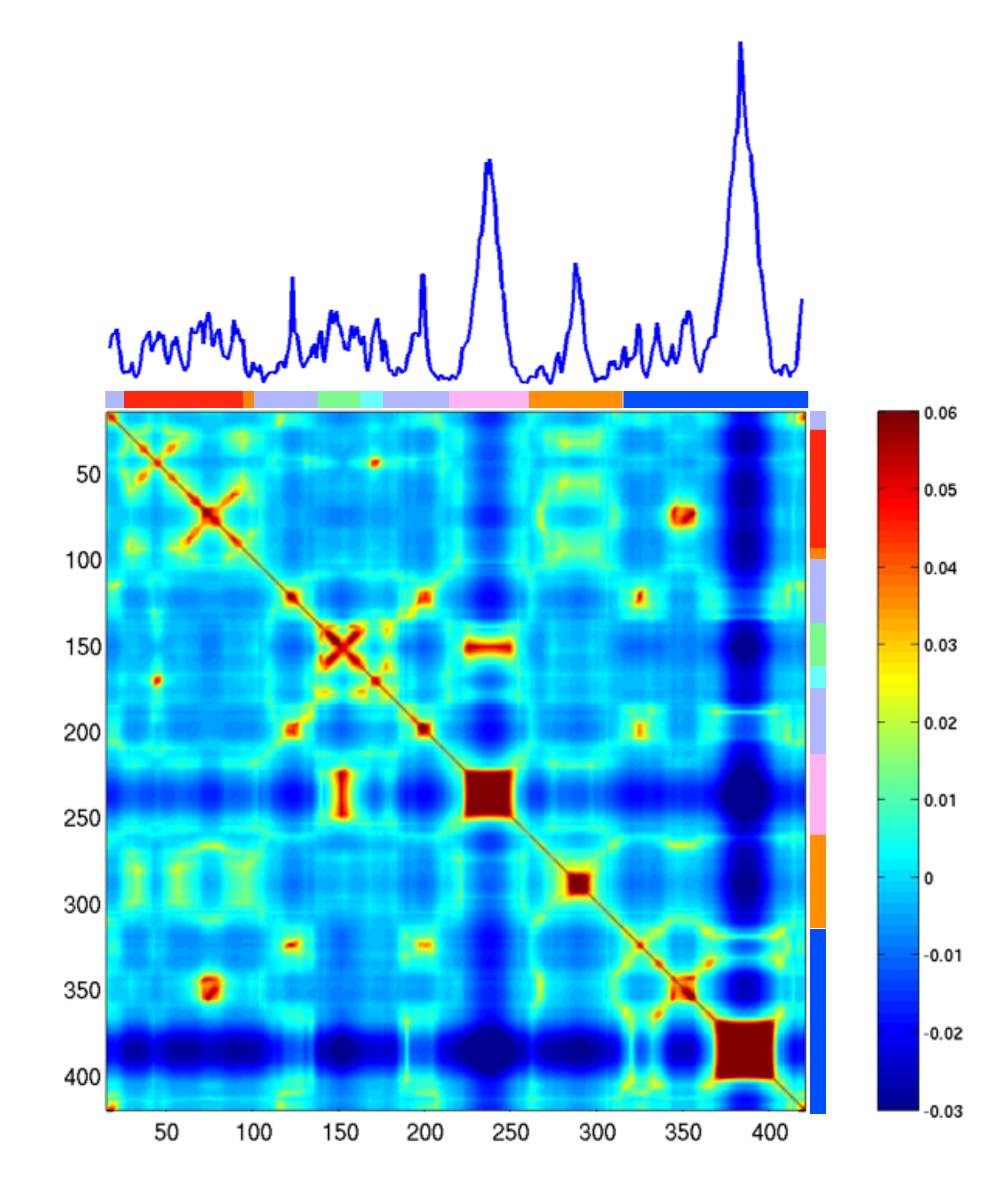}
	\caption{Dynamic cross-correlation (${\bf T}_{ij}$) of \emph{T}. ribozyme at $f=0$. 
	Mean square fluctuation is shown on the top. 
	}
	\label{fig:Tij}
\end{figure}

\begin{figure}
	\centering
	\includegraphics[width=1\linewidth]{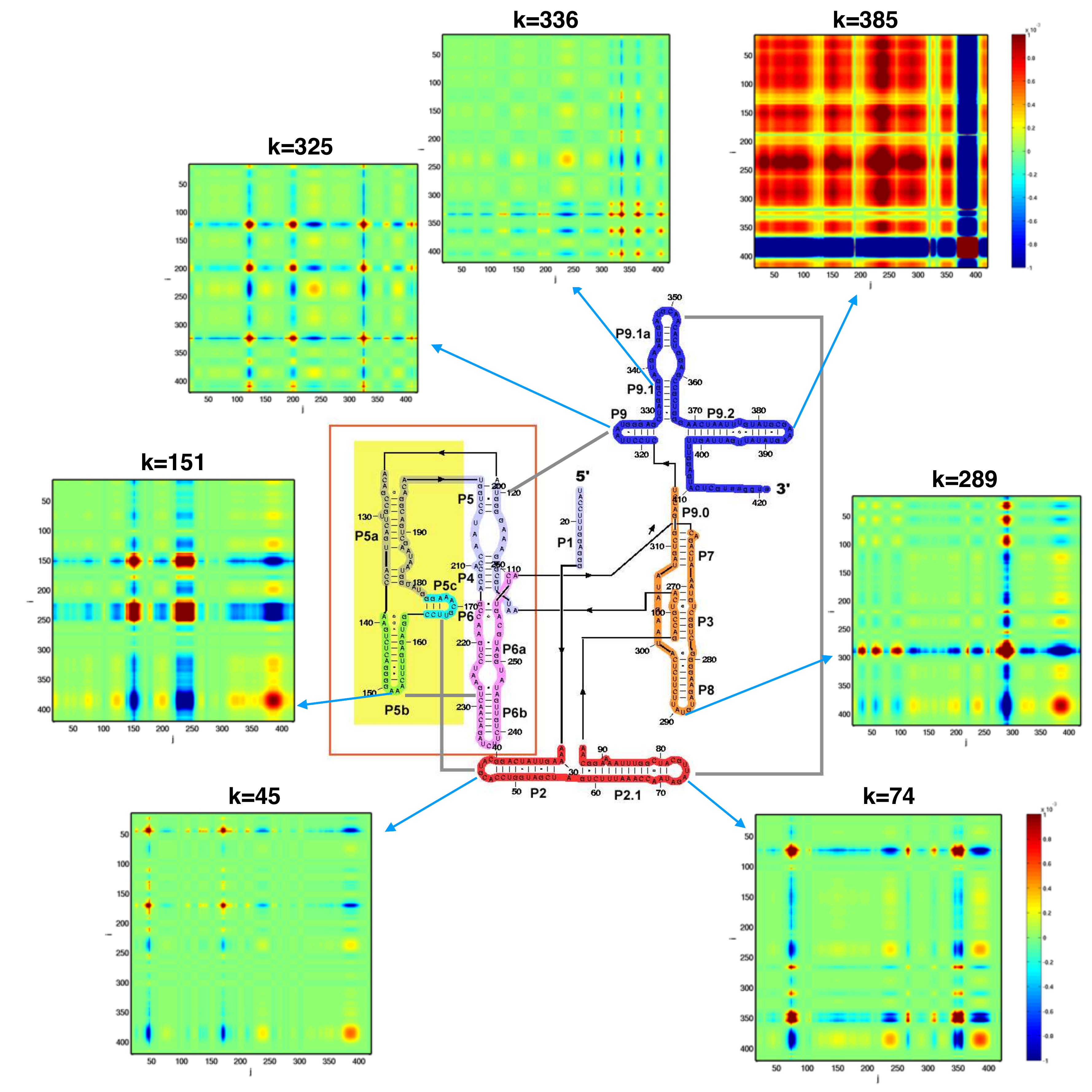}
	\caption{Allosteric communication over the \emph{T}  ribozyme structure when $k$-th residue is perturbed by mechanical force. 
	Cross-correlation matrices ${\bf F}^k$ are shown for seven different $k$ values.  Location  of each the perturbed site is marked with the arrow in the secondary structure.  
	All the results for $k=1,\ldots, N$ are available in the Supporting Movie. 
	}
	\label{fig:perturbation}
\end{figure}

To a first approximation, the conformational dynamics of biomolecules in response to the influence of the ambient temperature and mechanical perturbation is encoded in the three dimensional structure of the molecule \cite{Klimov00PNAS}. By using this finding we showed that the force response of RNA in the native state can be well described with GNM which incorporated the 3D  topology of the molecule in terms of Kirchhoff matrix. 
Importantly, the pattern of cross-correlation over the molecular architecture under mechanical perturbation differs from the thermal cross-correlation (Fig.\ref{fig:Tij}) and also shows a large variation depending on where the force is applied (Fig.\ref{fig:perturbation}, SI Movie). 

The accuracy of our prediction on dynamical cross-correlation ($\langle\delta{\bf R}_i\cdot\delta{\bf R}_j\rangle$) which is used to capture the essence of conformational dynamics is expected to increase with the growing system size ($N$) because main contributions to the value of $\langle\delta{\bf R}_i\cdot\delta{\bf R}_j\rangle$ come from low frequency modes. This is evident in the  mathematical structure, $\langle\delta{\bf R}_i\cdot\delta{\bf R}_j\rangle\sim \left(\sum_{k=2}^N\lambda_k^{-1}{\bf u}_k{\bf u}_k^T\right)_{ij}$ with $\lambda_2 < \lambda_3 < \ldots < \lambda_N$ \cite{BaharPRL97}. 

Finally, given the simplicity of the model, 
the extent of agreement of the predicted unfolding pathways from GNM with those identified by SM force measurement is surprising. 
We note that simulation of forced-unfolding process at the experimentally meaningful pulling speed using even the minimal SOP model is time-consuming for large RNA molecules.  
While the current model can be further improved to obtain  better agreement with experimentally determined properties, we rediscover that a number of key features of interest (conformational fluctuation, allostery, unfolding/folding pathways) are already encoded in the 3D native topology \cite{Klimov00PNAS,Hyeon06Structure}.   We should note, however, subtle features such as pathway bifurcation cannot obtained by GNM which only uses the contact map of the native state. Nevertheless, because of the simplicity the proposed method for RNA unfolding and the earlier study on proteins \cite{Srivastava13PRL} show that  approximate features of tension-induced rupture can be gleaned from the present model.
\\

{\bf Acknowledgement}
We are pleased to dedicate this work to Klaus Schulten who had the vision decades ago that computations could become an integral part of solving major problems in biology. His pioneering theoretical and computational studies is a testimony to that vision. We thank the KIAS Center for Advanced Computation for providing computing resources for this work. This work was supported in part by a grant (CHE 16-36424) from the National Science Foundation to DT.


\clearpage 

\end{document}